# Shifting of surface plasmon resonance due to electromagnetic coupling between graphene and Au nanoparticles

**Jing Niu,[1] Young Jun Shin,[1] Jaesung Son,[1] Youngbin Lee,[2] Jong-Hyun Ahn,[2] and Hyunsoo Yang[1,*]**

[1]*Department of Electrical and Computer Engineering, National University of Singapore, 117576, Singapore*
[2]*School of Advanced Materials Science & Engineering, SKKU Advanced Institute of Nanotechnology, Sungkyunkwan University, Suwon, 440-746, South Korea*
[*]*eleyang@nus.edu.sg*

**Abstract:** Shifting of the surface plasmon resonance wavelength induced by the variation of the thickness of insulating spacer between single layer graphene and Au nanoparticles is studied. The system demonstrates a blue-shift of 29 nm as the thickness of the spacer layer increases from 0 to 15 nm. This is due to the electromagnetic coupling between the localized surface plasmons excited in the nanoparticles and the graphene film. The strength of the coupling decays exponentially with a decay length of $d/R$=0.36, where $d$ is the spacer layer thickness and $R$ is the diameter of the Au nanoparticles. The result agrees qualitatively well with the plasmon ruler equation. Interestingly, a further increment of the spacer layer thickness induces a red-shift of 17 nm in the resonance wavelength and the shift saturates when the thickness of the spacer layer increases above 20 nm.

## 1. Introduction

Since graphene was first exfoliated from highly-oriented pyrolytic graphite in 2004, it has attracted enormous research interest because of its exceptional properties [1, 2]. Especially, it has demonstrated extraordinary electronic and thermal transport features as a promising material in the future nanoelectronics [3-5]. Its low opacity of ~ 2.3% over a wide frequency range, exceptional mechanical flexibility, and high electron mobility makes it an ideal material as transparent conductive electrodes [6-9]. Advances in large scale graphene synthesis methods enable the utilization of graphene as transparent conductive electrodes to successfully construct touch-screen panels and dye-sensitized solar cells [6, 10]. The true potential of graphene lies in photonics and optoelectronics [11]. Other demonstrated optoelectronic graphene devices include ultrafast lasers and broadband polarizers [12, 13]. In addition, taking advantage of the zero bandgap, short carrier lifetime, and high carrier mobility a graphene based photodetector in high speed communication links has been proposed [14]. One of the major drawbacks in the implementation of graphene optoelectronic devices is the low photocurrent generated by a single layer graphene sheet. In order to overcome this issue, metal nanoparticles were included in the system to enhance the photocurrent, and specific wavelength detection can be achieved by varying the structure of the metal nanoparticles simultaneously [15]. However, the presence of graphene near the metal nanoparticles consequently modifies the physical environment of the localized surface plasmon resonance (LSPR) excited in metal nanoparticles [16, 17]. Previously, a gold thin film was utilized with metal nanoparticles to achieve the tuning of the wavelength of LSPR by changing the distance between metal nanoparticles and the conductive film, which in turn varies the coupling strength of the electromagnetic field surrounding the particles and the conductive film [18].

In this work, we have investigated the coupling of the electromagnetic field between surface plasmons excited in gold nanoparticles and the anti-parallel image dipoles formed in graphene. The coupling strength of the field is controlled by inserting different thicknesses of an $Al_2O_3$ spacer layer between nanoparticles and graphene. As the spacer thickness increases from 0 to 15 nm, a blue-shift of the surface plasmon resonance from 599 to 570 nm is observed. This can be explained by the reduction of the coupling strength of the electromagnetic field of the excited plasmons in the nanoparticles and the anti-parallel image dipoles in graphene. The experimental results fit well with the plasmon ruler equation derived previously for the near-field electromagnetic field coupling [19]. The decay length is estimated to be 0.36. However, a further increment of the separation to 20 nm shifts the resonance wavelength back to a longer wavelength of 586 nm and the resonance wavelength saturates regardless of any further increment of the separation up to 35 nm. Our findings facilitate a better understanding of the electromagnetic coupling and provide an opportunity of wavelength selection in the graphene/spacer/nanoparticle system which could be utilized in multicolor selective optoelectronic devices.

## 2. Methods

Single layer graphene grown by chemical vapor deposition (CVD) on copper films is utilized in the experiment [6, 7]. The CVD graphene thin films are transferred to transparent borosilicate glass substrates for the transmission measurements. The quality of the graphene on borosilicate glass substrates is examined by Raman spectroscopy. As shown in Fig. 1(a), the absence of the D peak and a sharp 2D peak illustrates high-quality single layer graphene [20]. The transmission data in Fig. 1(b) without and with graphene show a difference of ~ 2.3%, which matches well with the opacity of single layer graphene [8]. A layer of Al thin film less than 3 nm is deposited on top of the graphene samples by electron beam evaporation, followed by natural oxidation under ambient conditions. In order to ensure that the Al thin film is fully oxidized into $Al_2O_3$, the above steps are repeated for a thicker $Al_2O_3$ film. The thickness of the film is monitored through quartz crystals during the deposition, and estimated by an ellipsometer after the oxidation process. The $Al_2O_3$ film functions as the spacer layer between graphene and metal nanoparticles. Subsequently, a 1.5 nm Au film is deposited to form Au nanoparticles. The structure of the sample is illustrated in Fig. 1(c). The size of the nanoparticles is examined by scanning electron microscope (SEM) as shown in Fig. 1(d), and there is no observable difference in nanoparticles for various thicknesses of $Al_2O_3$. An image processing software, ImageJ is utilized to analyze the size of the nanoparticles. The average diameter of the spherical nanoparticles is ~ 10 nm with a standard deviation of 2.4 nm. The variation of the LSPR wavelength in Au nanoparticles is carried out through transmission measurements in an UV-visible spectrophotometer. An unpolarized light source is used to excite LSPR on Au nanoparticles and the incident light illuminates the sample perpendicularly. The excitation of LSPR on Au nanoparticles causes extinction of the transmitted light. Therefore, a dip is observed in the transmission spectrum at the resonance wavelength.

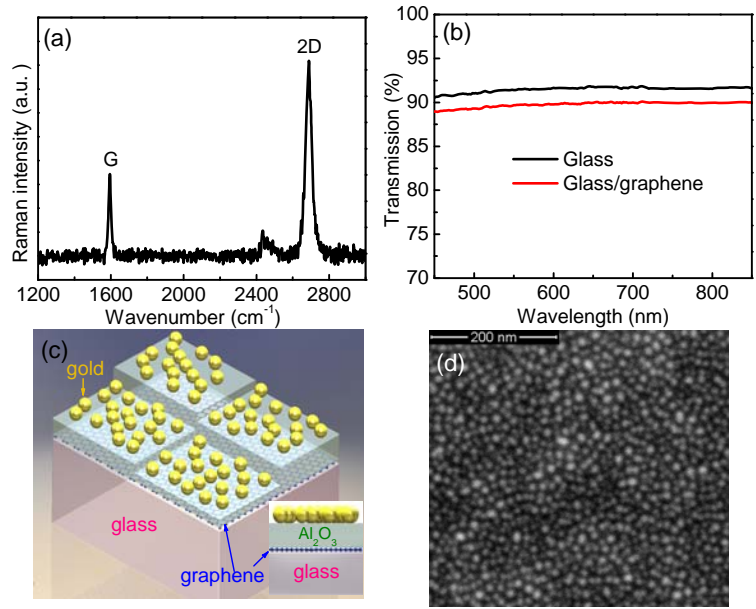

Fig. 1. (a) Raman spectrum of single layer CVD graphene with a 488 nm laser. (b) Transmission data of a borosilicate glass substrate without and with graphene. (c) Illustration of the sample structure (inset: cross section view of the device structure). (d) SEM image of Au nanoparticles formed on top of an $Al_2O_3$ spacer layer.

## 3. Results and discussion

Seven different $Al_2O_3$ films from 5 to 35 nm are deposited on bare glass substrates and graphene samples. The samples without graphene function as control samples. All samples have been processed together to minimize any experimental error due to fabrication condition changes. The transmission spectra of the samples have been measured right after the oxidation process. As shown in Fig. 2(a), the transmission spectra remain flat through the measurement range for glass samples capped with different thicknesses of $Al_2O_3$. The transmission difference is less than 3% between samples capped with various thicknesses of the $Al_2O_3$ film. A similar result is observed for graphene samples as shown in Fig. 2(b) with slightly smaller transmission values due to graphene. After the formation of Au nanoparticles on the samples, the transmission spectra are measured again. Figure 2(c) shows the transmission spectra of samples without graphene. The presence of transmission dips and its position agrees well with the resonance wavelength of Au nanoparticles, indicating the excitation of LSPR [21]. A large red-shift of the resonance is observed, when a 5 nm $Al_2O_3$ layer is introduced between the glass substrate and the Au particles compared to the case when there is no $Al_2O_3$ in the structure. A further increment of the thickness of the $Al_2O_3$ layer from 5 to 35 nm induces a small red-shift (~ 9 nm) of the LSPR. The transmission spectra of graphene samples with various $Al_2O_3$ thicknesses are shown in Fig. 2(d). Unlike the samples without graphene, a blue-shift of 29 nm is observed, when the thickness of the spacer layer increases from 0 to 15 nm. A further increment of the thickness to 20 nm causes a red-shift of the resonance wavelength and no further shifting is observed regardless of the increment of the spacer layer thickness.

Figure 3(a) shows the result of the theoretical calculation of the transmission value (1 − extinction efficiency) as a function of the separation between a gold nanosphere and a graphene substrate. The theoretical calculation is carried out based on dipole approximation, which is a common model to study the effect of a conductive film to the LSPR of metal nanoparticles [22]. The structure utilized in the calculation is shown in the inset of Fig. 3(a), in which a gold nanosphere is placed above a graphene substrate with a separation of *d*. The dielectric constant of graphene is calculated assuming that the optical response of a single graphene layer is given by the optical sheet conductivity, and the dielectric constant of gold is from the literatures [23, 24]. The calculated resonance wavelength as a function of the spacer layer thickness is shown in Fig. 3(b) as a blue line. A blue-shift of the resonance wavelength is clear for thinner insulating layers (0 to 10 nm), however, the resonance wavelength saturates when the thickness increases beyond 10 nm. This model correctly explains the observed blue-shift, but does not describe the subsequent red-shift nor does the model predict the correct resonance wavelength. Such deficiencies are presumably due to the assumptions of the model. For example, only one nanosphere is included in the calculation instead of many nanospheres. It has been observed that when nanoparticles are in close proximity, the coupling of the surface plasmon modes of nanoparticles will cause a red-shift of the resonance wavelength [25-28].

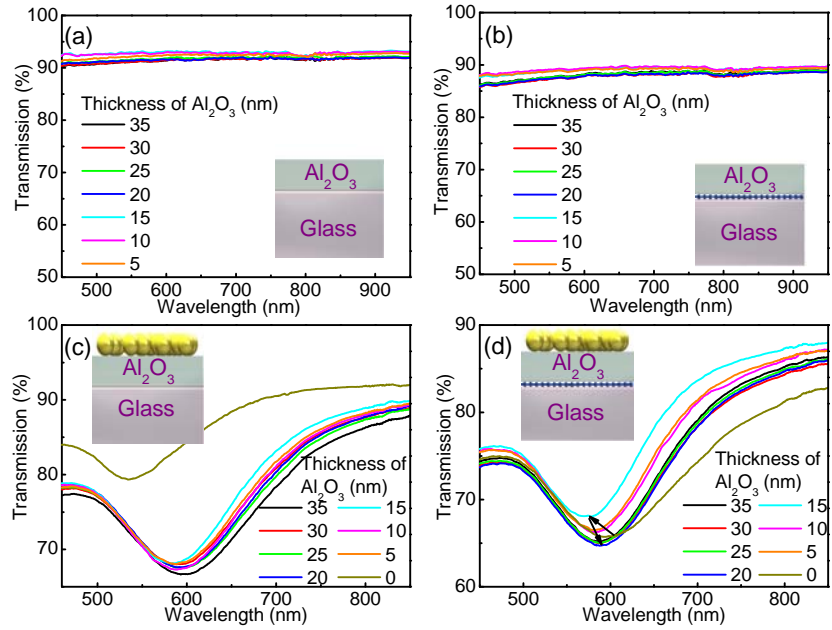

Fig. 2. (a) Transmission spectra of glass substrates capped with different thicknesses of $Al_2O_3$. (b) Transmission spectra from a structure of glass/graphene/$Al_2O_3$. (c) Transmission spectra from a structure of glass/$Al_2O_3$/particles. (d) Transmission spectra from a structure of glass/graphene/$Al_2O_3$/particles with various thicknesses of $Al_2O_3$. Each inset shows a cross section view of each sample structure.

The experimental and calculation results of the resonance wavelength as a function of the spacer layer thickness are summarized in Fig. 3(b). A different trend in the shifting of LSPR without and with graphene is obvious. For samples without graphene, a red shift of the LSPR can be explained by an increment of the relative permittivity of the physical environment, since the relative permittivity of $Al_2O_3$ is higher than that of glass and air [29, 30]. For samples with graphene, a blue-shift can be accounted for by the coupling of the electromagnetic field between the particles and the conducting film [22, 31]. When LSPR is excited in the nanoparticles, an anti-parallel image dipole of the resonance is induced in the metal film. A stronger electromagnetic coupling between the nanoparticles and the metal film causes a longer resonance wavelength. As the separation increases from 0 to 15 nm, the coupling strength reduces resulting in a blue-shift in the resonance wavelength. A fit of our experimental data for spacer layer thickness from 0 to 15 nm using an exponential equation is shown in Fig. 3(c). The plasmon ruler equation is given by $\Delta\lambda/\lambda_0 = a \times \exp(-x/\tau) + y_0$, where $a$, $\tau$ (decay length), and $y_0$ are the fitting parameters. $\lambda_0$ is the shortest resonance wavelength, $\Delta\lambda$ equals to the shifted wavelength compared to $\lambda_0$, $x$ is given by the spacer layer thickness (*d*) over the diameter of nanoparticles (*R*) [19, 32]. The equation is initially proposed for the particle-particle system to study the plasmon coupling in nanoparticles pairs, and has been also used for particle-substrate system, since the image dipole in the substrate/metal film can be regarded as the actual charge in the other particle [32]. The fitting of our experimental data shows a decay length of 0.36, which agrees well with the decay length (0.3) of the particle-substrate system reported previously [32].

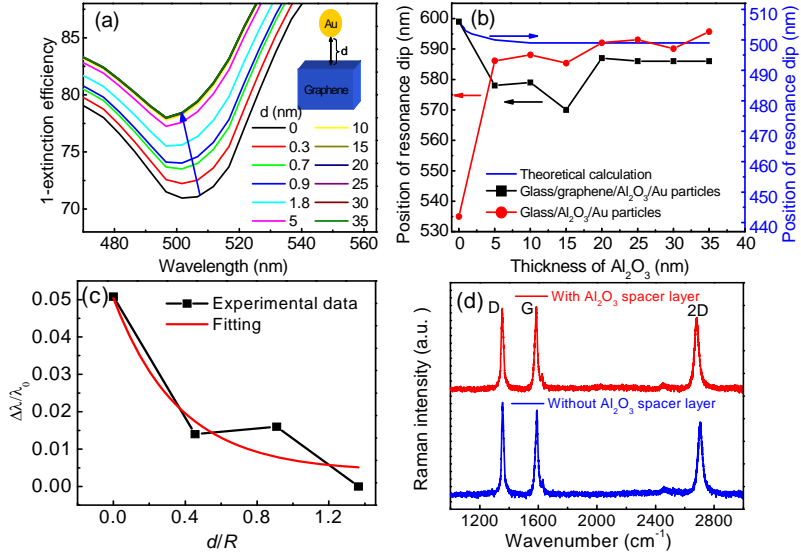

Fig. 3. (a) Calculation results of the LSPR wavelength excited by parallel electric fields (inset: structure used for calculation). (b) Dependence of the resonance wavelength on the spacer layer thickness for samples without and with graphene. (c) Fitting of experimental data with the plasmon ruler equation. (d) Raman spectra of samples after deposition processes.

A further increment of the distance between nanoparticles and graphene above 15 nm induces a red-shift and then a saturation of the resonance wavelength. The result is different from the previous studies of the two particles system and particle-substrate system, in which only blue-shift is observed with increasing of the spacer layer thickness [19, 32]. However, a similar effect has been reported for a silver nanoparticle and gold thin film system [18]. The blue-shift can be explained by the weakening of the coupling strength of the electromagnetic field as the distance between the particles and graphene increases. The reason of a red-shifting phenomenon could be very complicated, since for an intermediate spacer layer thickness, both the polarizability of the spacer layer and the charge response in the graphene film affect the LSPR of the metal nanoparticles [18]. In our theoretical model, the polarizability of the spacer layer is not considered. For large values of *d* the resonance wavelength is expected to saturate to the value of the system without graphene. As shown in Fig. 3(b), the resonance wavelength without and with graphene converges to a similar value for very thick spacer layers.

Raman spectra of the graphene samples after the deposition processes are measured as shown in Fig. 3(d) to evaluate the quality of graphene. For graphene capped with different thicknesses $Al_2O_3$, the spectra do not show any noticeable difference [16]. For a direct deposition of Au on top of graphene, a slightly smaller G to D peak ratio is observed. This is reasonable since the evaporation of Au is performed at a higher temperature compared to the case of Al. Although a D peak is present in the spectra, the G and 2D peaks are well preserved, demonstrating that the structural integrity of the graphene film is retained. The in-plane correlation length is ~ 4.3 nm and ~ 4.1 nm for graphene capped with and without spacer layer, respectively [33]. The in-plane correlation length is much larger than the conductivity lost limit of graphene [34, 35]. Therefore, graphene can still function well as a conductive layer.

## 4. Conclusion

In conclusion, by adjusting the thickness of the insulating spacer layer between Au nanoparticles and a single layer graphene thin film, the wavelength of LSPR can be tuned. As the separation between Au nanoparticles and graphene increases from 0 to 15 nm, the resonance wavelength has a blue-shift of approximately 29 nm. A further increment of the distance between these two parties causes a red-shift of the resonance, and the shifting saturates when the distance is more than 20 nm. The complex shifting behavior of the resonance wavelength can be understood by the electromagnetic coupling between graphene and particles, the relative permittivity of the surrounding media, and the polarizability of the spacer layer. Our study facilitates a comprehensive experimental study of the electromagnetic coupling of LSPR excited in Au nanoparticles and graphene. In addition, our finding suggests a straightforward and effective way of achieving multicolor selection in graphene/nanoparticles optoelectronic devices.

**Acknowledgements**

This work was supported by the Singapore National Research Foundation under CRP Award No. NRF-CRP 4-2008-06 and the National Research Foundation of Korea (2011-0006268).